\documentclass[aps,twocolumn,prb,superscriptaddress]{revtex4}
\usepackage[pdftex]{color,graphicx}
\usepackage{dcolumn}
\usepackage{amsmath}
\usepackage{natbib}

\begin{document}

\title{The role of the rare earth in the lattice and magnetic coupling in multiferroic h-HoMnO$_3$}

\author{J. Liu}
\author{Y. Gallais}
\author{M-A. Measson}
\author{A. Sacuto}
\affiliation{Laboratoire Mat\'eriaux et Ph\'enom\`enes Quantiques UMR 7162 CNRS, Universit\'e Paris Diderot-Paris 7, 75205 Paris cedex 13, France}
\author{S. W. Cheong}
\affiliation{Rutgers Center for Emergent Materials and Department of Physics and Astronomy, Rutgers University, 136 Frelinghuysen Road, Piscataway, NJ 08854, USA}
\author{M. Cazayous}
\affiliation{Laboratoire Mat\'eriaux et Ph\'enom\`enes Quantiques UMR 7162 CNRS, Universit\'e Paris Diderot-Paris 7, 75205 Paris cedex 13, France}

\date{\today}     

\begin{abstract}
We used Raman scattering to study the lattice and magnetic excitations in the hexagonal HoMnO$_3$ single crystals. 
The E$_2$ phonon mode at 237 cm$^{-1}$ is affected by the magnetic order. This mode is related to the displacement of Mn and O ions in a-b plane and modulates the 
Mn-O-Mn bond angles in a-b plane and the in-plane Mn-Mn superexchange interaction. 
The mode at 269 cm$^{-1}$ associated to the displacement of the apical Ho$^{3+}$ ions along the c direction presents an abrupt change of slope at T$_N$ showing that 
the role of the rare earth ions can not be neglected in the magnetic transition. 
We have identified magnon and crystal field excitations. The temperature dependence of the magnetic excitations has been compared to the Mn and Ho moment and indicates that the exchange interaction pattern
between Mn and Ho atoms drives the uniaxial anisotropy gap above the Mn-spin-rotation transition. 

\end{abstract}


\maketitle

\section{Introduction}

Multiferroics\cite{Eerenstein} which exhibit coexistence of magnetism and ferroelectricity with cross coupling are of great scientific and technological interest for further spintronics, magnonics and nanoelectronic devices. These include new forms of magnetoelectric memories, electrically-controlled magnonic elements or domain-wall-based devices with engineered magnetic structure.\cite{Allibe, Rovillain, kajiwara10, magnonics, Lee}
Additionally, these compounds can combine multiferroic orders and geometrical frustrations at low dimensionality that provide unusual spin dynamics as well as phase transitions.\cite{Kimura} 

Rare-earth manganites (RMnO$_3$) are one of canonical examples of multiferroics and are one of the most investigated family. These compounds crystallize in two different structures as a function of the ionic radius :
orthorhombic structures for larger ionic radius (R= La, Ce, Pr, Nd, Sm, Eu, Gd, Tb, Dy) and hexagonal structure (space group : P6$_3$cm ) for R with smaller ionic radius (R= Ho, Er, Tm, Yb, Lu, Y).\cite{Fiebig2005, Cheong2007, Park2003} Both exhibit multiferrocity but the microscopic origins are different. The magnetic frustrations in the orthorhombic manganites lead to spin-lattice coupling induced by the inverse Dzyaloshinski-Moriya interaction.\cite{Cheong2007}
The ferroelectricity in the hexagonal compounds results from electrostatic and size effects that lead to the bulking of MnO$_5$ bipyramids and the displacement of the rare-earth ions.\cite{Aken2004} 

For each rare-earth manganites, the interactions between the lattice and magnetic degrees of freedom need to be clarified. Hexagonal HoMnO$_3$ is one of the most studied of the rare-earth manganites and has
a magnetic field-temperature phase diagram that exhibits a multitude of complex spin structures \cite{Munoz2000, Fiebig2000, Fiebig2002}. Recently, the antiferromagnetic spin dynamics in HoMnO$_3$ has been studied tracking the changes in the magnon modes with Thz pulses.\cite{Bowlan2016} The development of the time resolved experiments to probe the phonon and spin dynamics show the importance to understand first the static interactions. Moreover, the stability of a magnetic configuration rather than an other and the nature of the spin waves have to be understood.
The magnetic properties of h-HoMnO$_3$ rare-earth manganite have been studied by inelastic neutron scattering\cite{Fabreges2009}, second harmonic generation\cite{Fiebig2000, Fiebig2002} and low heat transport\cite{Wang2010} whereas the lattice vibrations have been investigated by Raman and infrared spectroscopy.\cite{Litvinchuk2004} However, the studies of the static coupling between the lattice and magnetic degrees of freedom on HoMnO$_3$ single crystals remains small in number.\cite{Cruz2005, Poirier2011} 

Here, we have investigated the A$_1$ and $E_2$ phonon modes of h-HoMnO$_3$ with polarized Raman spectroscopy.  
The temperature dependence of particular phonon modes associated to motions modulating the super-superexchange paths between adjacent Mn planes and modulating the Mn-O-Ho interactions shows that the Mn-Mn and Ho-Mn interaction along the c-axis play an important role in the magnetic ordering in h-HoMnO$_3$. 
The magnetic excitations are also been measured, spin excitations have been identified. The temperature behavior of the magnon modes show that the interaction between the rare earth and the Mn atoms controls the uniaxial anisotropy gap from T$_N$ down to the Mn-spin-rotation transition. 

\par

\section{Experimental Details}

\par

HoMnO$_3$ single crystals were grown using the high-temperature flux growth technique in a platinum crucible.\cite{Kim2000}. The crystals are millimeter-size platelets with a thickness around 0.1 mm and a large surface of $0.5~cm^2$. The hexagonal c-axis is perpendicular to the surface.  The crystals have been polished to obtain high surface quality for optical measurements.    
Raman scattering is performed in backscattering geometry using the 532~nm excitation line from a laser diode and collected by a triple spectrometer Jobin Yvon T64000 equipped with a liquid-nitrogen-cooled CCD (charged-coupled device. The high rejection rate of the spectrometer allows to detect the low frequency excitations. Measurements between 10 and 300 K have been performed using an ARS closed-cycle He cryostat. 

\par

\section{Results and discussion}

\par

\begin{figure}
 \includegraphics*[width=9cm]{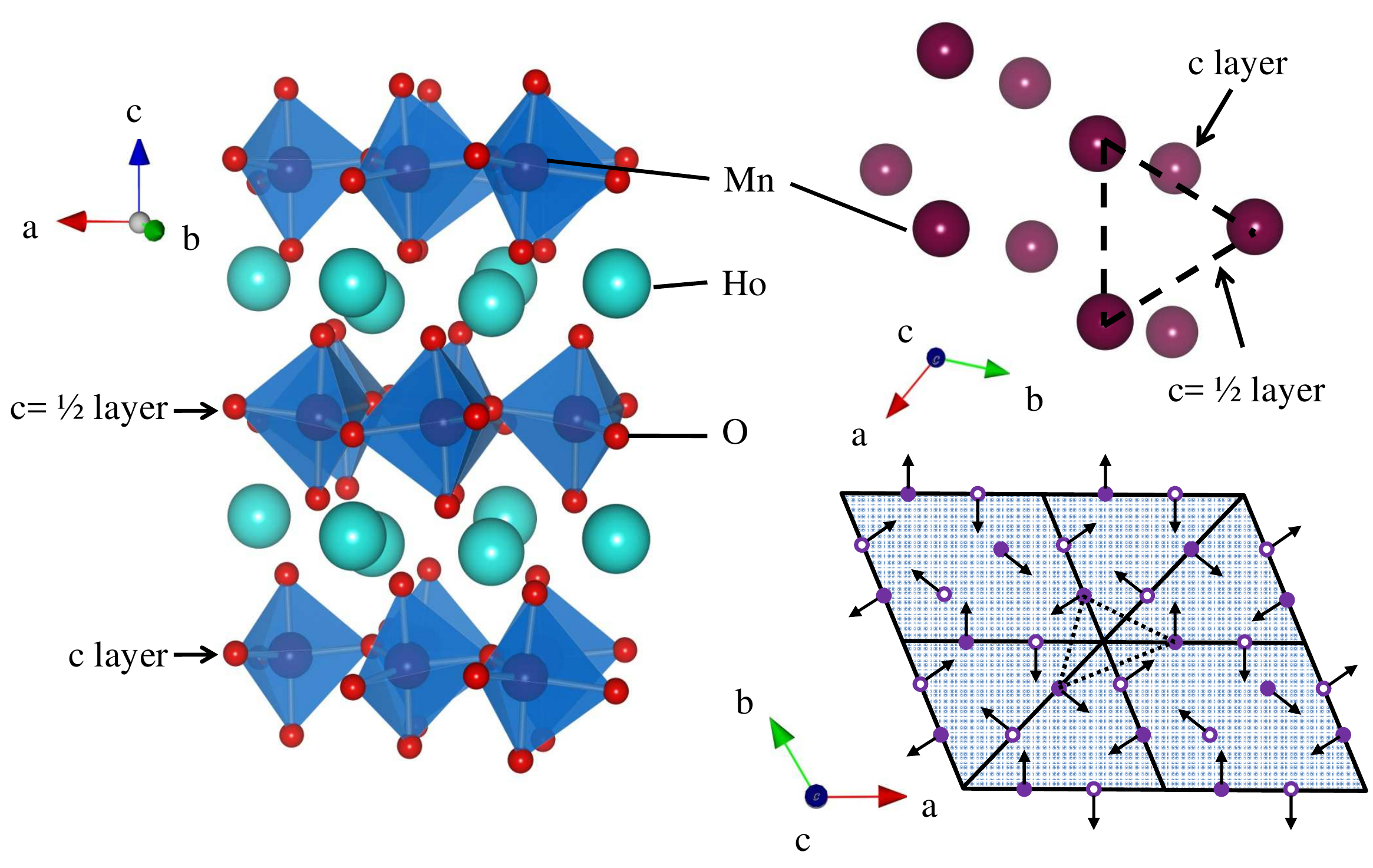} 
 \caption{\label{Fig1} 	 
 Crystallographic structure of hexagonal HoMnO$_3$ (P6$_3$cm phase) and the magnetic order.}
\end{figure}

HoMnO$_3$ crystallizes in a hexagonal lattice, space group P6$_3$cm. This compound is formed by layers of corner-sharing trigonal MnO$_5$ bipyramids arranged in a layered type structure in the a-b plane with apical (O$_1$, O$_2$) and in-plane (O$_3$, O$_4$) oxygen ions as shown in Fig. \ref{Fig1}.\cite{Lottermoser} Between the bipyramid layers, the rare-earth ions layers are stacked along c axis.
Ferroelectric polarization appears along the c axis below 875 K and results from electrostatic and size effects that lead to the buckling of MnO$_5$ bipyramids and the displacement of the Ho$^{3+}$ ions out of the (a,b) plane.\cite{Aken2004, Lottermoser} 
The moments of Mn$^{3+}$ ions display antiferromagnetic (AF) orderings at T$_{N,Mn}$=75K. The Mn$^{3+}$ ions form triangular planar sublattices and the AF exchange coupling among Mn$^{3+}$ moments is geometrically frustrated (see Fig. \ref{Fig1}). Below T$_{N,Mn}$, the Mn magnetic moments order in 120$^o$ arrangements.\cite{Vajk2005} Another transition occurs below 40 K and implies an in-plane Mn spin reorientation by 90$^o$.\cite{Munoz2000, Vajk2005, Fiebig2000}
Ho moments of the 4b sites are antiferromagnetically coupled within a given layer and Ho$^{3+}$ ions display AF orderings at T$_{N,Ho}$=4.6K. 
It has been proposed that the Ho$^{3+}$ moment are oriented along the $c$ axis with an Ising-type anisotropy. The Ho$^{3+}$ ions present in two different crystallographic sites form two sublattices that can order in parallel or antiparallel orientations.\cite{Brown, Hur}. The c and c=+1/2 layers are ferromagnetically coupled.  

\par

\subsection{Lattice excitations}

\par

The crystal symmetries define the matrix elements of the Raman tensor which are selected by the incident and scattered light polarizations. Thus selection rules define the vibrational modes that can be detected by Raman spectroscopy. 
The group-theoretical analysis shows that the zone-center phonon modes ($\Gamma$-point) are in the irreductible representations of the $C_{6v}$ : $10A_1+5A_2+10B_1+5B_2+15E_1+15E_2$. Only 38 of these modes are Raman-active: $\Gamma_{Raman} = 9A_1+14E_1+15E_2$.
Using different scattering configurations, it's possible to choose the mode to activate.\cite{Porto1966} Here we used incident wave vector anti-parallel to the scattered one (backscattering configuration).
The longitudinal optical A$_1$ modes are activated when the phonon propagation direction corresponds to the direction of the ions displacements using the z(xx)\={z} polarization configuration.
Pure E$_2$ modes are obtained using z(xy)\={z} geometry.

\begin{figure}
 \includegraphics*[width=9cm]{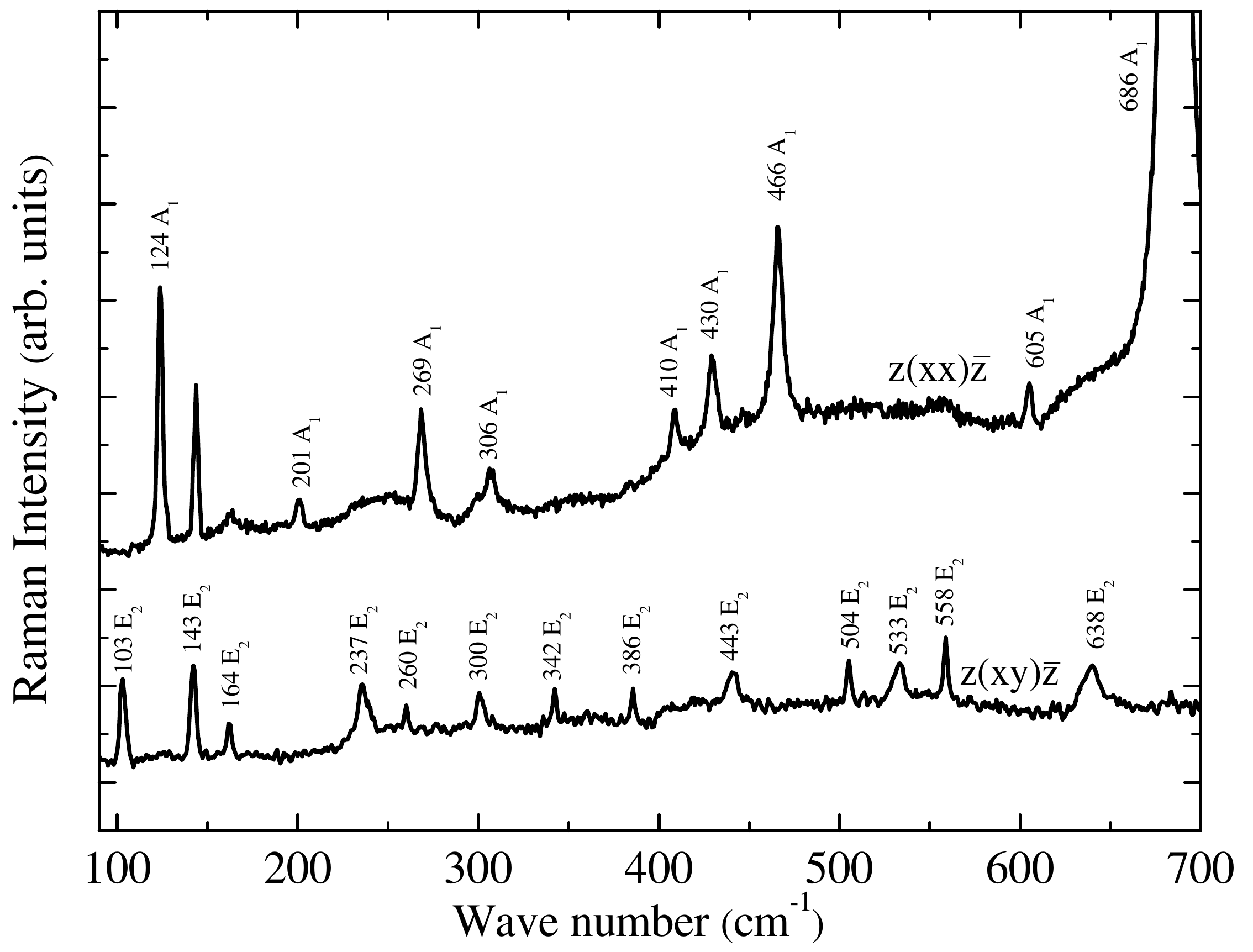} 
 \caption{\label{Fig2} 	 
Phonon modes in h-HoMnO$_3$ single crystal observed at 10 K using z(xx)\={z} (A$_1$ modes) and z(xy)\={z} (E$_2$ modes) scattering configurations.}
\end{figure}

Figure \ref{Fig2} presents the Raman spectra measured on h-HoMnO$_3$ single crystals with the z(xx)\={z} and z(xy)\={z} scattering configurations. The Raman selection rules allow us to identify the 9 A$_1$ modes and 13 E$_2$ modes over 15. 
The frequencies of the phonon modes at 10~K are reported in Table I and compared to the previous theoretical results on single crystal.\cite{Litvinchuk2004} 
We have measured the temperature dependences of the phonon modes in order to indentify the effect of the phase transitions. 

\begin{figure}
 \includegraphics*[width=8.5cm]{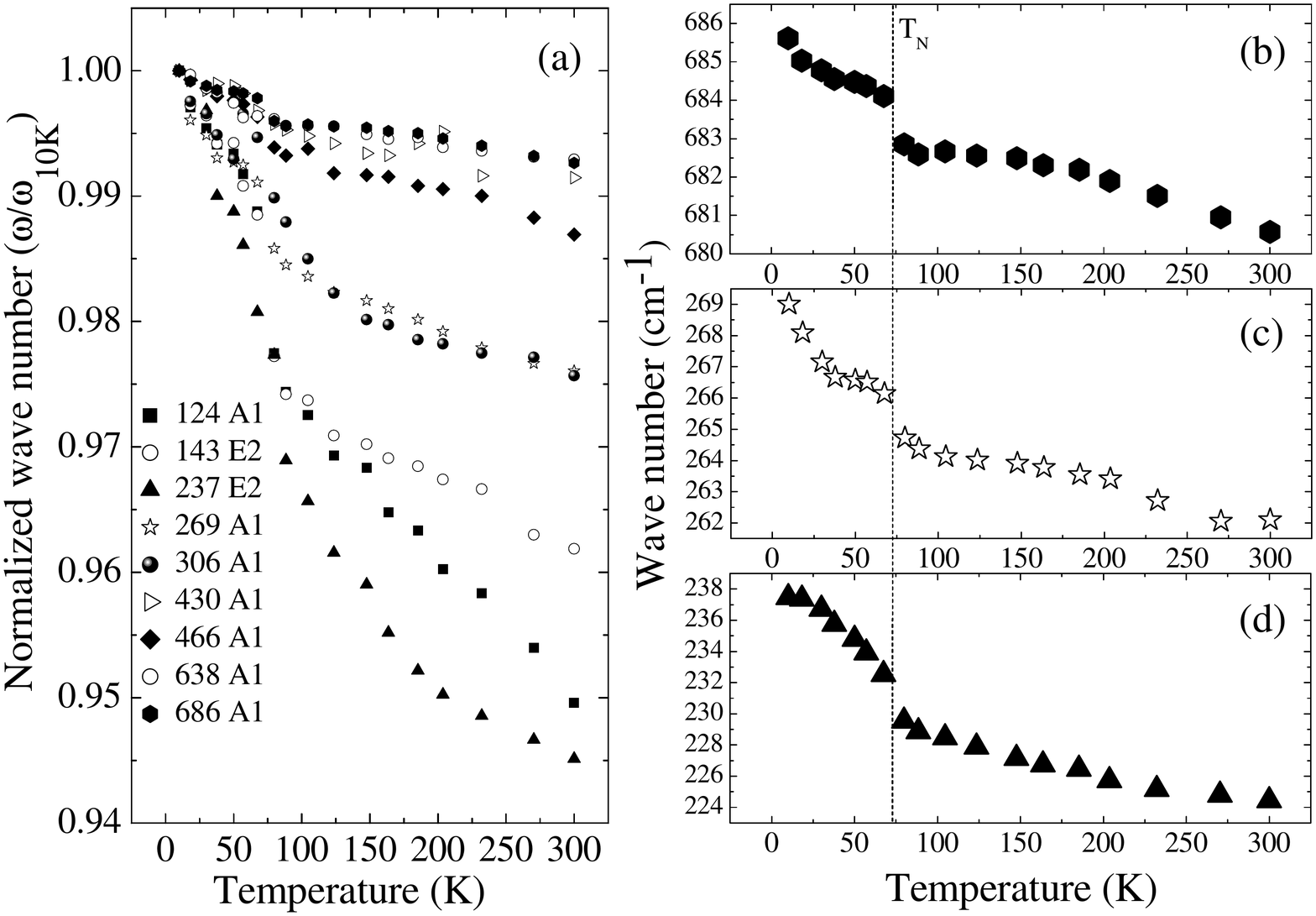} 
 \caption{\label{Fig3} 	 
a) Normalized wavenumbers ($\omega$(T)/$\omega$(10K)) of 7 A$_1$ and 2 E$_2$ modes and wavenumbers of the b) A$_1$ mode at 686 cm$^{-1}$, c) A$_1$ mode at 269 cm$^{-1}$ and d) E$_2$ mode at 237 cm$^{-1}$ as a function of temperature.}
\end{figure}

Figure \ref{Fig3}(a) presents the normalized frequencies over the frequency at 10 K of several A$_1$ and E$_2$ modes as a function of temperature.
All the phonon frequencies soften due to the dilation of the unit cell when temperature increases. Three phonon modes present an abrupt change in their frequencies as shown in Fig.~\ref{Fig3}(b, c, d) as the fingerprint of the spin-phonon coupling in the magnetically-ordered phase.

\begin{table}[ht]
  \begin{center}
  \parbox{8cm}
  \caption{ 
  
  A$_1$ and E$_2$ mode frequencies (cm$^{-1}$) measured in h-HoMnO$_3$ and description of the atomic displacements.}\\
	\vspace{3mm}
	 \begin{tabular}{ccccc}
	\hline\noalign{\smallskip}
	    Mode & This work & Theory\cite{Litvinchuk2004} & Direction of the \\
	       & & & largest displacement \\
	\hline\noalign{\smallskip}
	  A$_1$ & 124 & 127 & z(Ho)\\
	        & 201 & 234 & Rot x,y(MnO$_5$)\\
	        & 269 & 270 & +z(Ho)-z(Mn)\\
	        & 306 & 295 & x(Mn),z(O$_3$)\\
	        & 410 & 428 & +z(O$_3$,O$_4$)+x, y(O$_2$)-x,y(O$_1$)\\
	        & 430 & 460 & +z(O$_4$,O$_3$),-z(Mn)\\
	        & 466 & 474 & +x,y(O$_1$,O$_2$), -x,y(Mn)\\
	        & 605 & 614 &  +z(O$_1$,O$_2$)-z(Mn)\\
	        & 686 & 673 & +z(O$_1$)-Z(O$_2$)\\
  \hline\noalign{\smallskip}
	  E$_{2}$  & - & 64 & x,y(Ho,Mn)\\
	           & 103 & 96 & +x,y(Mn,O$_3$,O$_4$)-x,y(Ho)\\
	           & 143 & 137 & x,y(Ho)  \\
	           & 164 & 152 & x,y(Ho) \\
	           & 237 & 231 & +x,y(Mn),-x,y(O$_3$,O$_4$)\\
	           & - & 254 & z(Mn,O$_2$,O$_1$)\\
	           & 260 & 265 & z(Mn,O$_1$,O$_2$)\\
	           & 300 & 330 & z(O$_2$,O$_1$),x,y(O$_4$)\\
	           & 342 & 339 & +x,y(O$_1$,O$_2$,O$_3$,O$_4$)-x,y(Mn)\\
	           & 386 & 402 & +x,y(O$_1$,O$_4$)-x,y(O$_2$,Mn)\\
	           & 443 & 468 & +x,y(O$_4$)-x,y(O$_1$,Mn)\\
	           & 504 & 523 & x,y(O$_4$,O$_3$,O$_1$,O$_2$))\\
	           & 533 & 557 & x,y(O$_4$)\\
	           & 558 & 583 & x,y(O$_4$,O$_3$)\\
	           & 638 & 649 & x,y(O$_3$,O$_4$)\\
	  \hline\noalign{\smallskip}
	 \end{tabular}
	\end{center}
	\label{freqphonon}
 \end{table}

The E$_2$ mode at 237 cm$^{-1}$ presents a frequency shift beyond the mean behaviour of the other modes (Fig.~\ref{Fig3}(a)) and  an abrupt change of slope around the N\'eel temperature in Fig.~\ref{Fig3}(d). 
This phonon mode is related to the displacement of Mn and O ions in the a-b plane and it modulates the inplane Mn-Mn superexchange interaction and the Mn-O-Mn bond angles in a-b plane. 
Such a behaviour has been already measured in rare earth manganites like in YbMnO$_3$ single crystals.\cite{Liu2012} 
The peak at 269 cm$^{-1}$ and 686 cm$^{-1}$ show the same behaviour at the N\'eel temperature. 

The mode at 269 cm$^{-1}$ is associated to the relative displacement of the apical Ho$^{3+}$ ions along the c direction.
The role played by the R element in the RMnO$_3$ manganites has been underestimated. However the role of non magnetic R atom is not negligible as underlined by the temperature behaviour of the Ho$^{3+}$ ions frequency at the the N\'eel temperature.

The mode at 686 cm$^{-1}$ is related to the relative displacement of the apical oxygen ions along the c direction and modulates the Mn-O-O-Mn bond angles. Notice that the inter planar coupling between Mn cations and two O anions (Mn-O-O-M) corresponds to the inter planar super-superexchange paths.
The abrupt change of slope at T$_N$ shows that the super-superexchange path through the apical oxygen participates to the three dimensional magnetic ordering. 

\subsection{Spin excitations}

\begin{figure}
 \includegraphics*[width=9cm]{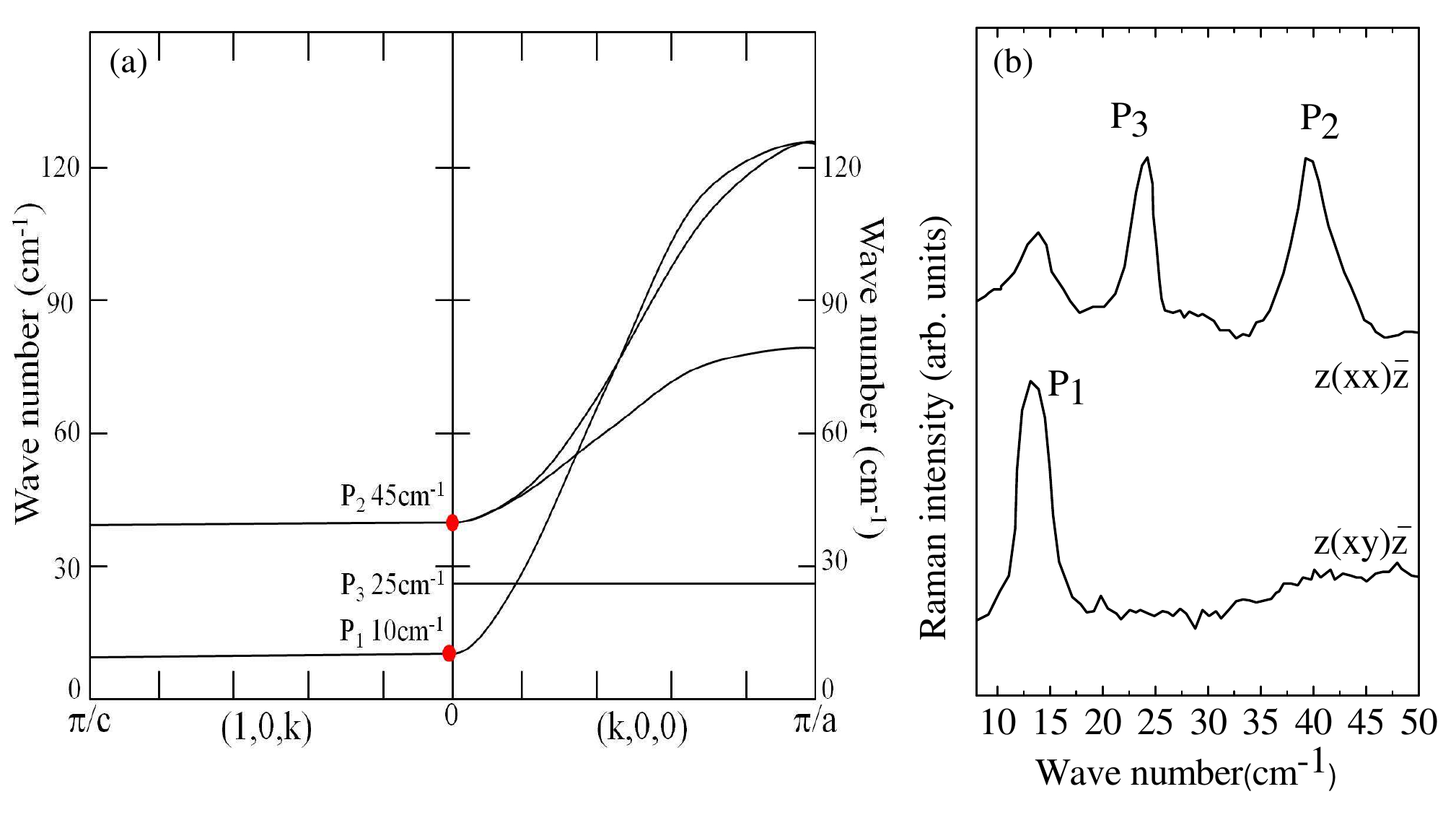} 
 \caption{\label{Fig4} 	 
a) Magnon dispersions along \textbf{k}= (1,0,k) and \textbf{k}= (k,0,0) extracted from Ref. \onlinecite{Fabreges2009}. b) Raman spectra of low frequency excitations measured at 10 K in z(xy)$\bar{z}$ and z(xy)$\bar{z}$) configuration.}
\end{figure}

Figure~\ref{Fig4}(b) shows the low frequency Raman spectra measured on h-HoMnO$_3$ single crystal at 10 K.
Three peaks are measured: P$_1$ = 13 cm$^{-1}$, P$_3$ = 25 cm$^{-1}$, P$_2$ = 41 cm$^{-1}$ with two polarization configurations.
Remember that Raman scattering can probe in addition to phonon modes, the crystal field, one magnon (zone center) and two-magnon (zone edge) excitations. 
Below 50 cm$^{-1}$, there is no phonon modes, and as shown in the following paragraph the  P$_1$, P$_2$ and P$_3$ peaks disappear at the N\'eel temperature, so these peaks can be thus attributed to magnetic excitations. 
In order to determine the origin of the magnetic excitations, we have compared our measurements to neutron scattering results. 
Figure \ref{Fig4}(a) represents the magnetic excitation dispersions along  the \textbf{k}= (1,0,k) and \textbf{k}= (k,0,0) axes.\cite{Fabreges2009}
At the $\Gamma$-point, the mode at 10 cm$^{-1}$ corresponds to the uniaxial anisotropy gap and can be associated to the P$_1$ peak. 
The mode at 45 cm$^{-1}$ in Fig.\ref{Fig4}(a) is splitted at higher energies at the zone-edge with a higher branch around 130 cm$^{-1}$ and a lower branch branch around 80 cm$^{-1}$. 
The P$_2$ peak can be attributed to this mode and corresponds to the one-magnon mode of the Mn$^{3+}$ magnetic structure in the (a,b) plane, in particular to the global in phase and out of phase rotations of the 120$^o$ pattern inside the basal plane.
The P$_3$ is related to the crystal field excitation at 25 cm$^{-1}$ which presents a flat dispersion as shown by neutron scattering.\cite{Fabreges2009} Transition from the ground state of the 4f ions to excited state are in principle Raman active and correspond to crystal field excitations. 

\begin{figure}
 \includegraphics*[width=9cm]{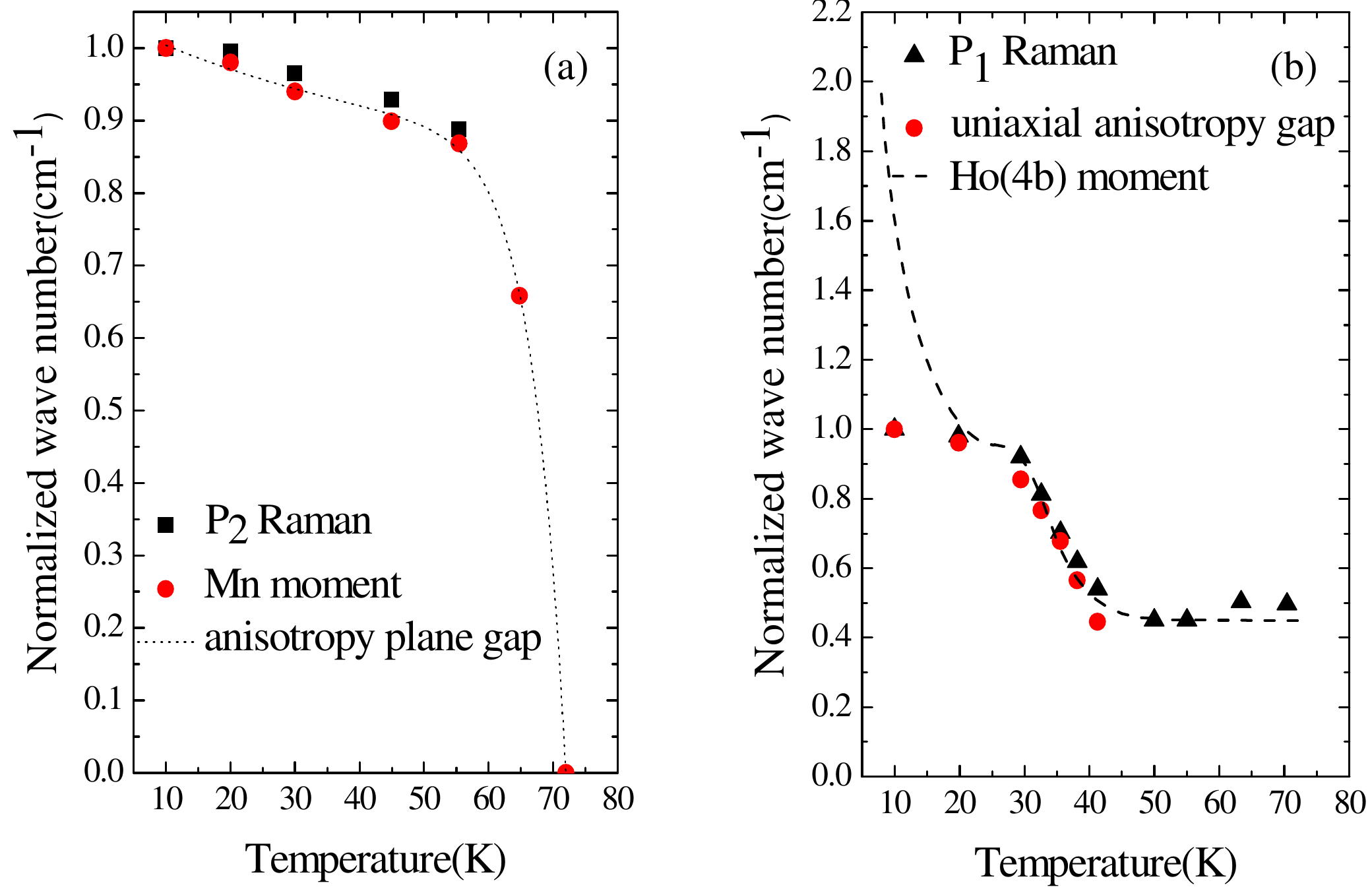} 
 \caption{\label{Fig5} 	 
Temperature dependence of normalized wavenumber ($\omega$(T)/$\omega$(10K)) of a) the magnetic mode P$_2$ compared with the Mn moments and the energy of the anisotropy plane gap\cite{Fabreges2010}, and of b) the P$_1$ mode compared to the uniaxial energy gap and the Ho(4b) moment.\cite{Fabreges2010}}
\end{figure}

The temperature dependence of the magnon wavenumber P$_1$ and P$_2$ recorded from 10 to 70 K is shown in Fig. \ref{Fig5}.

In Fig. \ref{Fig5} (a), the Raman shift of the P$_2$ mode decreases as the temperature is increased. This conventional softening is first compared to the one of the Mn moment. The good agreement with the variation of the Mn moment shows that this magnon mode can be thus associated to the spin excitations of the Mn$^{3+}$ ions. It also agrees well with the temperature dependence of the anisotropy plane gap measured by neutron scattering.\cite{Fabreges2010} The P$_2$ peak is the signature of the anisotropy planar gap in the (a,b) plane and varies with temperature similar to the Mn$^{3+}$ ion moment. 

In Fig. \ref{Fig5} (b), the P$_1$ peak wave number presents a change in the slope between 30 and 40 K. We have compared this behaviour to the thermal evolution of the uniaxial gap.\cite{Fabreges2010} The good agreement confirms that this peak is related to the uniaxial anisotropy gap. At T$_{SR}$=33 K a Mn-spin-rotation transition has been observed implying an in-plane Mn-spin reorientation by 90$^{o}$ and an antiferromagetic order of the Ho moments along the c axis. To shed some light on the role of the Ho ions, we have drawn in Fig. \ref{Fig5} (b) the temperature dependence of Ho moments. The P$_1$ peak frequency shift is close to the behaviour of the Ho moment down to T$_{SR}$. This evidence points out the role of the Ho$^{3+}$ ion moment and of the Ho-Mn interaction on the P$_1$ spin wave. 
Below T$_{SR}$, the uniaxial gap is no more correlated to the Ho moment.
Those measurements show that the exchange interaction pattern between the rare earth and Mn atoms is a key to understand the uniaxial anisotropy gap but not sufficient below Mn-spin-rotation transition.\\   

\par

\section{conclusion}

\par

In summary, our measurements show that the E$_2$ mode  associated to the relative displacement of the apical Ho$^{3+}$ ions along the c direction is sensitive to the the N\'eel temperature pointing out the non negligible role played by the R element in the RMnO$_3$ manganites. We also reveal the spin excitations and the role played by the Ho$^{3+}$ ion moment and of the Ho-Mn interaction on the uniaxial anisotropy gap. 



\begin{thebibliography}{00}

\bibitem{Eerenstein}	W. Eerenstein, N. D. Mathur, and J. F. Scott, Nature {\bf 442}, 759 (2006).

\bibitem{Allibe} J. Allibe, S. Fusil, K. Bouzehouane, C. Daumont, D. Sando, E. Jacquet, C. Deranlot, M. Bibes, and A. Barth\'el\'emy, Nano Lett. {\bf 12}, 1141 (2012).

\bibitem{Rovillain} P. Rovillain, R. De Sousa, Y. Gallais, A. Sacuto, M. M\'easson, D. Colson, A. Forget, M. Bibes, A. Barth\'el\'emy, M. Cazayous, Nat. Mater. {\bf 9} 975 (2010).
	
\bibitem{kajiwara10} Y. Kajiwara, K. Harii, S. Takahashi, J. Ohe1, K. Uchida, M. Mizuguchi, H. Umezawa, H. Kawai, K. Ando, K. Takanashi, S. Maekawa, and E. Saitoh, Nature {\bf 464}, 262 (2010).

\bibitem{magnonics} S.O. Demokritov, and A.N. Slavin, Springer-Verlag, Berlin, 2013. 

\bibitem{Lee} J. H. Lee, I. Fina, X. Marti, Y. H. Kim, D. Hesse, and M. Alexe, Adv. Mater. {\bf 26}, 7078 (2014).

\bibitem{Kimura} T. Kimura, T. Goto, H. Shintani, K. Ishizaka, T. Arima, and Y. Tokura, Nature {\bf 426}, 55 (2003).

\bibitem{Fiebig2005} M. Fiebig, J. Phys. D {\bf 38}, R123-R150 (2005).

\bibitem{Cheong2007} S. W. Cheong and M. Mostovoy, Nature Mater. {\bf 6}, 13 (2007).

\bibitem{Park2003} J. Park, J. G. Park, G. S. Jeon, H. Y. Choi, C. H. Lee, W. Jo, R. Bewley, K. A. McEwen, and T. G. Perring, Phys. Rev. B {\bf 68}, 104426 (2003).

\bibitem{Aken2004} B. B. Van Aken, T. T. M. Palstra, A. Filippetti, and N. A. Spaldin, Nature Mater. {\bf 3}, 164 (2004).

\bibitem{Munoz2000} A. Mu$\tilde{n}$oz, J. A. Alonso, M. J. Martinez-Lope, M. T. Casais, J. L. Martinez, and M. T. Fernandez-Diaz, Chem. Mater. {\bf 13}, 1497 (2001).

\bibitem{Fiebig2000} M. Fiebig, D. Fr$\ddot{o}$hlich, K. Kohn, S. Leute, Th. Lottermoser, V. V. Pavlov, and R. V. Pisarev, Phys. Rev. Lett. {\bf 84}, 5620 (2000).

\bibitem{Fiebig2002} M. Fiebig, C. Degenhardt, and R. V. Pisarev, J. Appl. Phys. {\bf 91}, 8867 (2002).

\bibitem{Bowlan2016} P. Bowlan, S. A. Trugman, J. Bowlan, J. X. Zhu, N. J. Hur, A. J. Taylor, D. A. Yarotski, and R. P. Prasankumar, Phys. Rev. B {\bf 94}, 100404 (2016).

\bibitem{Fabreges2009} X. Fabr$\grave{e}$ges, S. Petit, I. Mirebeau, S. Pailhes, L. Pinsard, A. Forget, M. T. Fernandez-Diaz, and F. Porcher,  Phys. Rev. Lett. {\bf 103}, 067204 (2009).

\bibitem{Litvinchuk2004} A. P. Litvinchuk, M. N. Iliev, V. N. Popov and M. M. Gospodinov,  J. Phys. : Condens. Matter {\bf 16}, 809 (2004).

\bibitem{Cruz2005} C. dela Cruz, F. Yen, B. Lorenz, Y. Q. Wang, Y. Y. Sun, M. M. M. Gospodinov, and C. W. Chu, Phys. Rev. B {\bf 71}, 060407(R) (2005).

\bibitem{Wang2010} X. M. Wang, C. Fan, Z. Y. Zhao, W. Tao, X. G. Liu, W. P. Ke, X. Zhao, and X. F. Sun, Phys. Rev. B {\bf 82}, 094405 (2010).

\bibitem{Poirier2011} M. Poirier, J. C. Lemyre, P. O. Lahaie, L. Pinsard-Gaudart, and A. Revcolevschi, Phys. Rev. B {\bf 83}, 054418 (2011).

\bibitem{Kim2000} T. Choi, Y. Horibe,	H. T. Yi,	Y. J. Choi,	Weida Wu and S.-W. Cheong, Nature Mater. {\bf 9}, 253 (2010).

\bibitem{Lottermoser} T. Lottermoser, T. Lonkai, U. Amann, D. Hohlwein, J. Ihringer, and M. Fiebig, Nature {\bf 430}, 541 (2004).

\bibitem{Vajk2005} O. P. Vajk, M. Kenzelmann, J. W. Lynn, S. B. Kim, and S-W. Cheong,  Phys. Rev. Lett. {\bf 94} 087601 (2005).

\bibitem{Brown} P. J. Brown, and T. Chatterji, Phys. Rev. B {\bf 77} 104407 (2008).

\bibitem{Hur} N. Hur, I. K. Jeong, M. F. Hundley, S. B. Kim, and S.W. Cheong, Phys. Rev. B {\bf 79}, 134120 (2009).

\bibitem{Porto1966} S. P. S. Porto, J. A. Giordmaine, and T. C. Damen, Phys. Rev. {\bf 147}, 608 (1966).

\bibitem{Liu2012} J. Liu, C. Toulouse, P. Rovillain, M. Cazayous, Y. Gallais, M-A. Measson, N. Lee, S. W. Cheong, and A. Sacuto, Phys. Rev. B {\bf 86}, 184410 (2012).

\bibitem{Fabreges2010} X. Fabr$\grave{e}$ges, PhD Thesis, Etude des propriétés magnétiques et du couplage spin/réseau dans les composés multiferroïques RMnO$_3$ hexagonaux par diffusion de neutrons, University Paris-Sud 11, France (2010).
































\end{thebibliography}
\end{document}